\documentstyle[aps,graphics]{revtex}

\def\ph{\overline{\phi}}
\def\fm{f_{\rm{m}}}

\begin{document}
\draft
\title{Phase transitions in a ferrofluid at magnetic field induced
microphase separation}
\author{D. Lacoste and T. C. Lubensky}
\address{Department of Physics,
University of Pennsylvania, Philadelphia, PA 19104-6396, USA}
\date{\today}
\maketitle
\begin{abstract}
In the presence of a magnetic field applied perpendicular
to a thin sample layer, a suspension of magnetic colloidal
particles (ferrofluid) can form spatially modulated phases
with a characteristic length determined by the competition
between dipolar forces and short-range forces opposing
density variations.  We introduce models for thin-film
ferrofluids in which magnetization and particle density are
viewed as independent variables and in which the non-magnetic
properties of the colloidal particles are described either by
a lattice-gas entropy or by the Carnahan-Starling free
energy.  Our description is particularly well suited to the
low-particle density regions studied in many experiments.
Within mean-field theory, we find isotropic, hexagonal and
stripe phases, separated in general by first-order phase
boundaries.
\end{abstract}

\section{Introduction}
\label{sec:intro}
Ferrofluids are suspensions of ferromagnetic particles
 with a diameter of
about $10$nm in a carrier fluid. The particles are stabilized against
aggregation by coating with polymers for oily ferrofluids or with
charged surfactant for aqueous ferrofluids. On macroscopic scales, ferrofluids
can be described as super-paramagnetic liquids \cite{rosensweig}. The
application of a magnetic field perpendicular to a thin layer induces
microphase separation in a homogeneous aqueous or oily ferrofluid with
no surfactant and leads to the formation of a periodic lattice of
unbranched \cite{bacri1,hong} or branched \cite{wang} concentrated
phase columns. In thin layers of ferrofluid confined together with
an immiscible non-magnetic liquid, the columns can merge into sheets
\cite{flament}, and at a higher field, the sheets evolve into a
disordered
labyrinthine structure \cite{rosensweig}. In pure ferrofluids and in
ferrofluid emulsions, only the hexagonal phase of columns has been
reported \cite{bibette}. Similar periodic structures have been observed
in other physical systems such as Langmuir monolayers \cite{andelman},
magnetic garnet thin films \cite{garel}, or type I superconductors
\cite{faber}, as discussed by Seul {\it et al.} \cite{seul}. In all these
systems, there is a spontaneous spatial modulation of an order parameter, which
can be either the concentration or the magnetization of the particles,
or a combination thereof. The period of the modulation is determined by
the competition between long-range
 dipolar forces and short-range forces
favoring constant density. It depends on the magnetic field and the
thickness of the sample layer as discussed in a recent study on the
aggregate size and spacing formed in a thin film of ferrofluid
\cite{ytreberg}.

This paper concerns the thermodynamic stability and pattern
formation in a suspension of ferromagnetic particles in a
carrier fluid in the presence of a magnetic field applied
perpendicular to the sample layer. We formulate models for
thin films of these suspensions in which particle
concentration and magnetization, determined by the degree of
alignment of magnetic moments as well as particle
concentration, are treated as independent variables.   We
discuss two possible models for the entropy of the fluid, the
lattice gas model and the Carnahan-Starling model
\cite{carnahan}. Although the Carnahan-Starling model has
already been used for magnetic fluids \cite{morozov}, it has
not been used to describe transitions from a disordered to an
ordered phase in ferrofluids.  Our lattice model is
essentially identical to that of Sano {\it et al.} \cite{doi}
except that we consider the complete wave-number dependence
of interactions rather than the infinite wavenumber limit
appropriate to needle-like magnetic domains. Several previous
studies, including those of Andelman {\it et
al.}\cite{andelman} and Cebers\cite{cebers,cebers2}, are
based on Landau expansion of a lattice-gas model in the
vicinity of the liquid-gas critical point present in the
absence of the long-range part of the dipolar interactions.
The free energies of these
models are even-order expansion up to fourth order in the
deviation of the local density (which includes the modulated
and spatially uniform components) from the liquid-gas
critical density.  Our model free energy is in principle
valid for arbitrary values of the spatially
uniform part of the particle density. We place
no restrictions on the spatially uniform components of the
density, but we do expand the free energy in a power series
in the spatially varying component of the density.  The
entropy of the lattice-gas model is invariant under $(\phi -
\case{1}{2})\rightarrow -(\phi - \case{1}{2})$, where $\phi$
is the volume fraction of ferrofluid particles.  The
Carnahan-Starling entropy possesses no such symmetry. As a
result, its phase diagram, as we shall see, is more
asymmetric than that of the lattice-gas model, and its
interesting features occur at $\phi<1/2$.  We investigate the
phase diagrams of our model within mean-field theory in which
modulated hexagonal and stripe phases are described by sine
waves with wave numbers of a fixed magnitude.  Our results
are in qualitative agreement with those of
Cebers\cite{cebers,cebers2} and of Halsey\cite{halsey} but
differ from them in detail, particularly in the low-density
regions in which non-magnetic interactions are best described
by the Carnahan-Starling free energy.

\section{Helmoltz Free Energy}
\label{sec:model}

The magnetic particles are assumed to have a spherical ferromagnetic
core
of radius $a$, coated by a sheath of surfactant $\delta$. Due to steric
hindrance, the particles cannot come closer than a distance
$d=2(a+\delta)$. Each particle of ferrofluid is a magnetic
single-domain
of magnetic moment
\begin{equation}
 m_0=\frac{4\pi}{3}a^{3}M_{s}\mu_{0},
\end{equation}
where $M_s$ is the saturation magnetization of the bulk material, and
$\mu_0$ is the magnetic permeability of the vacuum. The volume of the
particles is $v_0=\pi d^3 /6$. It is assumed that the suspension is
monodisperse and that each particle carries the same magnetic moment.
We define $m({\mathbf{r}})$ to be the ratio of the average
magnetic moment divided by $m_0$, so that $0 \leq m \leq 1$,
 and $\phi({\mathbf{r}})$ to be the volume fraction of the
ferrofluid at point ${\mathbf{r}}$.
We treat the particles as hard spheres, and we
include only magnetic dipolar interaction beyond hard sphere repulsion.

The total free energy $F$ of the ferrofluid in a magnetic
field breaks up into four main contributions: the free energy
of independent magnetized particles in a magnetic field
$F_{\rm{m}}(m,\phi)$, the dipolar interaction energy
$E_{\rm{dip}}(m,\phi)$, the entropic contribution of
the hard spheres fluid, and the energy
cost associated with deviations of $\phi$  from spatial
uniformity $F_{\rm{nu}}(\phi)=k_B T LA  \int
d^2{\mathbf{r}}\left(\nabla \phi\right)^2 /(2v_0)$. In this
last term, $L$ is the thickness of the slab, and  $A$ is a
parameter with units of (length)$^2$, independent of the
magnetic field, which is related to the structure  factor at
low scattering angle \cite{lubensky}.

We have used two different forms of the entropy: the entropy of a gas
on a lattice, which has the following form per site
\begin{equation}
\label{eq:entropy}
s(\phi)=\phi \log \phi + (1-\phi) \log (1-\phi),
\end{equation}
and the entropy,
\begin{equation}
\label{eq:entropyCS} s(\phi)=\phi \left( \log \phi + \phi
\frac{4-3\phi}{(-1+\phi)^2} \right),
\end{equation}
of a Carnahan-Startling fluid\cite{carnahan}. In the
absence of further interactions, the free energies $-k_{B}T s(\phi
)$ of these models are convex, and their equilibrium
stable phase is a single-phase fluid with spatially uniform
$\phi$. As can be seen from Eq.~(\ref{eq:entropy}) and
Eq.~(\ref{eq:entropyCS}), the entropy of the lattice model is
an even function of $\phi-0.5$, whereas the entropy of the
Carnahan-Starling fluid does not have this symmetry. The
function $F_{\rm{m}}$ follows from a Langevin approach
(see appendix \ref{a:fm} for the derivation of the function $\fm(m)$):
\begin{equation}
\label{eq:fm0}
\frac{F_{\rm{m}}}{k_B T}=\frac{L}{v_0}\int d^2{\mathbf{r}}\phi
\left(\fm(m)-mh\right),
\end{equation} with
\begin{equation}
\label{eq:fm}
\fm(m)=m {\mathcal{L}}^{-1} (m)
 -\log\left(\frac{\sinh \left({\mathcal{L}}^{-1}(m)\right)}
{{\mathcal{L}}^{-1}(m)}\right),
\end{equation}
with ${\mathcal{L}}^{-1}$ denoting the inverse of the Langevin
function. The second term in Eq.~(\ref{eq:fm0}) is
the energy of the dipoles in the magnetic field $h$, and the
first term, the function $\fm(m)$ represents
the rotational entropy of the dipoles.
In Eq.~(\ref{eq:fm0}), we have introduced
\begin{equation}
\label{eq:defh}
h=\frac{Hm_0}{k_B T}
\end{equation}
which is a unitless measure of the
external magnetic field $H$.

The dipolar interaction energy (in SI units) can be written generally as:
\begin{equation}
\label{eq:Edip} E_{\rm{dip}}=\frac{{m_0}^2}{8 \pi \mu_0} \sum_{\alpha,\beta}
 m_{\alpha}^{i} m_{\beta}^{j} \left( -\nabla_{i}\nabla_{j}
   \frac{1}{|{\mathbf{r}}_{\alpha}-{\mathbf{r}}_{\beta}|} \right),
\end{equation}
where $i$ and $j$ are Cartesian coordinates, and $m_{\alpha}$
(respectively $m_{\beta}$) is the dimensionless magnetic
 moment of the particle located
at the point ${\mathbf{r}}_{\alpha}$ (respectively
${\mathbf{r}}_{\beta}$). Since ${\mathbf{m}}_{\alpha}$ represents an angular
average of the dipole moment, it is directed along the
$z$-axis, which is taken to be the direction of the applied
magnetic field. In the following, we will assume that
${\mathbf{m}}_{\alpha}=m_{\alpha}{\mathbf{e}}_z$ is independent
 of the $z$ coordinate. In a
continuous description of the medium, which will be discussed in the
next section, the local magnetization is
\begin{equation}
\label{eq:magnetization}
M({\mathbf{r}})=m_0 m({\mathbf{r}}) \frac{\phi
({\mathbf{r}})}{v_0},
\end{equation}
where $m({\mathbf{r}})$ is the coarse-grained unitless magnetic
moment at ${\mathbf{r}}$.
Inserting this equation into the continuum limit of Eq.~(\ref{eq:Edip}) gives
\begin{equation}
\label{eq:Edip2}
\frac{E_{\rm{dip}}}{k_B T}=\frac{\tilde{\lambda} L}{2v_0} \int
d^{2}{\mathbf{r}}d^{2}{\mathbf{r}}'
\phi({\mathbf{r}})
 \phi({\mathbf{r}}')m({\mathbf{r}})m({\mathbf{r}}')g({\mathbf{r}},
{\mathbf{r}}'),
\end{equation}
where $\tilde{\lambda}$ is a measure of the dipole-dipole interaction:
\begin{equation}
\tilde{\lambda}=24 \cdot \lambda, \, \, \, \, \, \, \,
{\rm with} \, \, \, \, \,
\lambda=\frac{m_0^2}{4\pi \mu_0 d^3 k_B T}
=\frac{\mu_0 M_s^2 4 \pi a^6}{9 d^3 k_B T},
\end{equation}
where $\lambda$ is the parameter introduced by Rosensweig
\cite{rosensweig}. The 2D Fourier
transform of the function $g({\mathbf{r}},{\mathbf{r}}')$
present in Eq.~(\ref{eq:Edip2}), is defined as
\begin{equation}
g({\mathbf{r}},{\mathbf{r}}')=\int \frac{d^2 {\mathbf{q}}}{(2\pi)^2}
g({\mathbf{q}}) e^{i\pi {\mathbf{q}} \cdot
({\mathbf{r}}-{\mathbf{r}}') }.
\end{equation}
It depends only on $q=|\mathbf{q}|$,
and it takes the form
\begin{equation}
\label{eq:g}
g(q)=\frac{1}{qL}\left(1-\exp(-qL)\right)-\frac{1}{3},
\end{equation}
which can be interpreted as the dipolar part of the pair
correlation function \cite{degennes,garel}. The first term in
Eq.~(\ref{eq:g}), is the long range contribution of the
interaction, which tends to $1$, the demagnetizing factor
of a film, as $q\rightarrow 0$, and to $0$ the demagnetizing
factor of a needle as $q\rightarrow \infty$. The second term
is the short range contribution, due to the local field
induced by the surrounding magnetic dipoles. In this geometry
where the applied magnetic field is perpendicular to the
sample layer, the dipoles are parallel to each other and
perpendicular to the plane the layer. The attraction between
dipoles in the head-to-tail configuration no longer appears
in Eq.~(\ref{eq:g}) because of the integration over the
thickness of the sample implicit in the derivation of
Eq.~(\ref{eq:Edip2}).  Note that $g(q=0) = 2/3$ is
positive.  This means that the total free energy is always
stable with respect to spatially uniform fluctuation in
$\phi$.  When $q\rightarrow \infty$, $g(q)$ tends to the
limit $-1/3$ determined entirely by local fields.  It is
the fact that $g(q)$ becomes negative for $qL$ greater than a
critical value that makes any transition from the spatially
uniform state possible.  Where it not for the negative local
field term, there would be no equilibrium spatially modulated
phases.

Finally the Gibbs free energy $F$ (for a chemical potential $\mu$) can
be expressed in terms of dimensionless lengths using the length
$\sqrt{A}$ and the transformations ${\bf r} \rightarrow \sqrt{A} {\bf
r}$ and $q\rightarrow q/\sqrt{A}$. The resulting dimensionless free energy
$f=(F v_0)/(k_B T A L)$, is
\begin{eqnarray}
\label{eq:gibbs}
\nonumber
f=\int d^{2} {\mathbf{r}}
\left[ \phi ({\mathbf{r}})
\left(\fm(m({\mathbf{r}}))-m({\mathbf{r}})h -\mu \right)
+ s(\phi({\mathbf{r}}))
+  \frac{1}{2} \left(\nabla \phi({\mathbf{r}})\right)^2  \right] \\
+ 12\lambda \int d^{2} {\mathbf{r}} d^{2}{\mathbf{r}'}
g({\mathbf{r}},{\mathbf{r}'})  \phi({\mathbf{r}})
\phi({\mathbf{r}}')m({\mathbf{r}})m({\mathbf{r}}') ,
\end{eqnarray}
where the Fourier transform of $g({\bf r}, {\bf r}')$ is
given by Eq.\ (\ref{eq:g}) with $q$ replaced by $q/\sqrt{A}$.
Thus, $g(q)$ as a function of the dimensionless $q$ is a
function of $q l$, where $l = L/\sqrt{A}$.

\section{Determination of the Phase Diagram}
\label{sec:result}
We first look at a spatially uniform state of the ferrofluid. This
state of the ferrofluid corresponds to a minimum of the total free
energy $f$ at $m=\overline{m}$ and $\phi=\ph$. It is obtained from
the equations
\begin{equation}
{\partial f \over \partial m} = 0 \qquad
{\rm and} \qquad {\partial f \over \partial \phi} = 0 .
\end{equation}
The first of these equations yields
\begin{equation}
\label{eq:equil}
\overline{m}={\mathcal{L}}(h_e),  \, \, \, \, \,
{\rm with} \, \, \, \, \, h_e= h-24 \lambda g(0) \ph\overline{m}.
\end{equation}
The second equation determining $\ph$ must in general be solved
numerically. It can however be solved exactly for the lattice gas
model. The result is
\begin{equation}
\label{eq:equil_lat}
\ph=\frac{1}{1+\exp(-\mu)h_e/\sinh (h_e) }.
\end{equation}
Equations (\ref{eq:equil},\ref{eq:equil_lat})
reproduce the main results of the
model of Sano {\it et al.} \cite{doi} and Cebers \cite{cebers}
for a spatially uniform ferrofluid,
when $g(0)$ is replaced by $g(q\rightarrow
\infty)$ in Eq.~(\ref{eq:equil}).  In our problem, however,
$g(0)$ is always positive so that there is no instability
toward the formation of coexisting homogeneous phases.

Spatially non-uniform configurations can be studied by expanding the
free energy difference $\Delta
f(\phi,m)=f(\phi,m)-f(\ph,\overline{m})$
in powers of $\delta m=m({\mathbf{r}})-\overline{m}$
 and $\delta \phi=\phi({\mathbf{r}})-\ph$.
The quadratic part of this free energy
 difference $\Delta f_{quad}$ has a simple form in
terms of the 2D Fourier transform $\delta m({\mathbf{q}})$ and
$\delta \phi({\mathbf{q}})$
\begin{equation}
\label{eq:df}
\Delta f_{quad}=\int \frac{d^2{\mathbf{q}}}{(2\pi)^2}
\left[ \frac{1}{2}r_{11}|\delta m({\mathbf{q}})|^2
+ \frac{1}{2}r_{22}|\delta \phi({\mathbf{q}})|^2
+ r_{12} \delta m({\mathbf{q}})  \delta \phi({\mathbf{-q}}) \right],
\end{equation}
with
\begin{equation}
\label{eq:det}
r= \left( \begin{array}{cc}
24 \lambda g(q)\ph^2+\fm''(\overline{m})\ph & 24 \lambda g(q)\overline{m}\ph \\
24 \lambda g(q)\overline{m}\ph  & 24 \lambda g(q)\overline{m}^2+s''(\ph)+q^2
\end{array}
\right).
\end{equation}
The coefficient $r_{22}$ is the most important term
determining at what values of $\lambda$ transitions occur,
and it is worth investigating it in more detail.  As already
discussed, $g(q)$ decreases monotonically from $2/3$ to
$-1/3$ as $q$ increases from $0$ to $\infty$. Since $q^2$
grows monotonically with $q$, $24\lambda g(q) {\overline
m}^2 + q^2$ has a minimum at $q=q^*$.  When $q^*l \gg 1$,
$g(q)\sim 1/ql-1/3$, and $q^*$ can be evaluated analytically:
\begin{equation}
\label{eq:qmin}
q^* = \left[ \frac{12\lambda}{l}{\mathcal{L}}^2 (h_e) \right]^{1/3}.
\end{equation}
The only negative term in $r_{22}$ is $-\lambda/3$
coming from the local field term in $g(q)$.  Thus, we can
write in general that
\begin{equation}
r_{22} (q^*) = 8 \overline{m}^2 (\lambda_c - \lambda ) ,
\end{equation}
where
\begin{equation}
\label{eq:lambdac}
\lambda_c = \frac{1}{8{\overline m}^2}
\left( 3 q^{*2} + s''(\ph) \right),
\end{equation} when $q^* l \gg
1$.  From this, we can see that the system becomes unstable
to the formation of modulated phases as $\lambda$ grows.

The spatially uniform phase becomes globally unstable with
respect to the formation of modulated phase when the
determinant, $J(q) = r_{11} ( q ) r_{22} ( q ) - r_{12}^2 ( q
)$, of the matrix in Eq.~(\ref{eq:det}) evaluated at its
minimum over $q$ becomes negative.  A significant
simplification of the theory results when fluctuations in $m$
are effectively frozen out, which occurs at ${\overline m}
=1$ when $\fm''({\overline m}) = \infty$.  In this ideal limit,
only $\phi$ varies spatially in modulated states, and the
stability of the uniform state is determined entirely by
$r_{22} ( q )$ rather than by $J(q)$.
As long as $r_{11}(q) >0$, an effective theory in terms of $\phi_q$
alone can be obtained by integrating over fluctuations in
$m$.  Removing $m$ will lead to
renormalization of the coefficients of $(\delta \phi)^n$ for
all $n$, as shown in appendix \ref{a:eff}.
The term with $n=2$ is the most important for the
determination of the phase diagram.  Its value in the
effective theory is
\begin{equation}
\label{eq:renorm}
r(q) = r_{22} ( q ) - {r_{12}^2(q) \over r_{11}(q)}
\approx r_{22} ( q ) - {(24 \lambda g(q) \overline{m})^2 \ph
\over \fm''( {\overline m})},
\end{equation}
where the later form is valid provided that
$\fm''(\overline{m}) \gg 8 \lambda \ph$.
The second term in this equation leads to a small shift in
$q^*$, the most unstable wavenumber from its value determined
by $r_{22}(q)$ alone  [Eq.\ (\ref{eq:qmin}) for $q^*l \gg 1$]
and to a small shift in $\lambda_c$, the critical value of
$\lambda$.
These shifts are small when $\overline{m}$ is sufficiently close
to $1$. For instance, for $\lambda=0.578$, $l=1000$ and
$\ph=0.5$, the second term is much smaller than the
first by at least two orders of magnitude for $h>14.2$,
which corresponds to $\overline{m}>0.9$. This is verified
in particular at the critical field for the transition to the modulated phases,
which occurs in this case at $h=21.5$ and $\overline{m}=0.94$.

For more general values of $\lambda$ however, $r_{11}$ can
in fact become negative. Unlike $r_{22}(q)$, $r_{11}(q)$ has no
stabilizing $q^2$ term in the current theory. As a result, it
reaches its minimum value of $(\fm''(\overline{m})-
8\lambda \ph)\ph$ at $q=\infty$.
Thus if $\fm''(\overline{m})< 8\lambda \ph$,
$r_{11}$ is negative for a range of $q$ and $J(q)$ will be
negative for $r_{22}>0$. This would indicate an instability toward
a phase with very short wavelength modulations. Since our theory
does not treat short wavelength physics in detail, we will
consider only situations in which $r_{11}$ is positive. In this
case, an effective theory in terms of $\phi_q$ alone can be
constructed by integrating over fluctuations in $m$.

In constructing phase diagrams in the $h-\ph$ plane,
we have used the two models discussed above: model $1$ in which
the fluctuations in $m$ are ignored, and model $2$ in which they
are included. Both models treat only terms up to fourth order in
$\phi_q$. Our theory in terms of $\phi$ only is very much in the
spirit of the single order parameter theory of Cebers
\cite{cebers,cebers2}, except for the shifts discussed above in $q^*$
and $\lambda_c$, which arise when fluctuations in $m$ are
included.
We assume for simplicity that
modulated phases are described by Fourier components with
reciprocal lattice vectors of the smallest possible magnitude.
Higher Fourier components undoubtedly can become important
particularly near the hexagonal-to-stripe transition. The effects of
these higher Fourier components, which are best treated using real
space ansatzes, will be treated elsewhere.
The free energy of the different phases are
\begin{itemize}
\item The isotropic phase with Helmoltz free energy
\begin{equation}
\label{fiso}
f_{\rm iso}=-12 \lambda g(0)\ph^2 \overline{m}^2
- \log \left( \frac{\sinh h_e}{h_e} \right) \ph  +s(\ph),
\end{equation}
\item The stripe phase
with a free energy
\begin{equation}
\label{fs}
f_s=f_{\rm iso} + \frac{1}{4}r \phi_{q}^2+
u_s \phi_{q}^4,
\end{equation}
In model 1, the stripe phase corresponds to a modulation
$\delta \phi({\mathbf{r}})=\phi_{q} \cos(q^* y)$, and the coefficients in
Eq.~(\ref{fs}) are $r=r_{22}(q^*)$ and
$u_s=s^{(4)}(\ph)/64$. In model 2, the stripe phase is characterized
by $\delta \phi({\mathbf{r}})=\phi_{q} \cos(q^* y)$ and $\delta
m({\mathbf{r}})=m_{q} \cos(q^* y)$. The coefficients $r=r(h,\ph)$ of
Eq.~(\ref{eq:renorm}) and $u_s$
of Eq.~(\ref{eq:us}) evaluated at $q^*$ must be used.
\item The hexagonal phase with a free energy
\begin{equation}
\label{fhex}
f_{\rm hex}=f_{\rm iso} + \frac{3}{4}r \phi_{q}^2+
 v \phi_{q}^3+u_h \phi_{q}^4,
\end{equation}
In model 1, the hexagonal phase corresponds to a modulation
$\displaystyle{\delta \phi({\mathbf{r}})=\sum_{i=1}^3 \phi_{q}
  \cos({\mathbf{q}}_i \cdot {\mathbf{r}} +\delta_i )}$
with $|{\mathbf{q}}_i|=q^*$ and $\displaystyle{\sum_{i=1}^3
 {\mathbf{q}}_i} =0$.
The coefficients in Eq.~(\ref{fhex}) are $r=r_{22}(q^*)$,
$v=s^{(3)}(\ph)/4$ and $u_h=15 s^{(4)}(\ph)/64$.
In model 2, the hexagonal phase corresponds to a
 modulation of this type for both
$\delta \phi$ and $\delta m$, and the coefficients
$r=r(q^*)$, $v$ and $u_h$ are given by
Eqs.~(\ref{eq:renorm}),(\ref{eq:v}) and (\ref{eq:u_h}).
\end{itemize}

A few comments about the general properties of our models are useful.
When $v(h,\ph)$ is zero, the
free energy of the stripe phase is  lower than that of the
hexagonal phase as can be seen by minimizing Eqs.~(\ref{fs})
and (\ref{fhex}) over $\phi_q$.  The energy density of the
stripe phase at its minimum over $\phi_q$ tends to zero
as $r^2$ as $r \rightarrow 0$ for $r<0$.  Thus, at
$\ph=\ph_c$ and $h=h_c$, determined by $v(h_c,\ph_c)=0$ and
$r(h_c,\ph_c)=0$, there is a second-order mean-field
transition from the isotropic to the stripe
phase \cite{brazovskii}.  When $v(h,\ph)$ becomes nonzero, the
hexagonal phase has lower free energy than the stripe phase
at small but nonzero $r$.  Thus, there will in general be a
transition from the isotropic to the hexagonal phase away from
$\ph_c$, and the isotropic, stripe, hexagonal, and
stripe phases will meet at the point $\ph=\ph_c$, $h=h_c$.
This is indeed the topology obtained in previous
calculations\cite{cebers,cebers2,halsey}.  The lattice-gas
entropy is invariant under $\psi = (\phi -
\case{1}{2})\rightarrow - \psi$, and it, therefore, only has
even order terms in a power series expansion in $\psi$ about
$\psi =0$. Thus $s^{(3)} ( \ph = \case{1}{2}) = 0$, and the
critical point in model 1 will occur at $\ph_c = 1/2$. The
Carnahan-Starling entropy has no reflection symmetry, and
in model 1 $s^{(3)}(\ph_c) = 0$ at $\ph_c = 0.1304$.
The non-entropic terms in
the total free energy are not invariant under $\psi
\rightarrow - \psi$. The result is a slight asymmetry in the
phase diagram for the
Carnahan-Starling model about $\ph_c$.
In model 2, $\ph_c$ and $h_c$ can only be determined by the
numerical solution of $v(h_c,\ph_c)=0$ and
$r(h_c,\ph_c)=0$. In our calculations, we find that this solution
is in general close to the value obtained for model 1.

\begin{figure}
{\par\centering
\resizebox*{8cm}{8cm}{\rotatebox{0}{\includegraphics{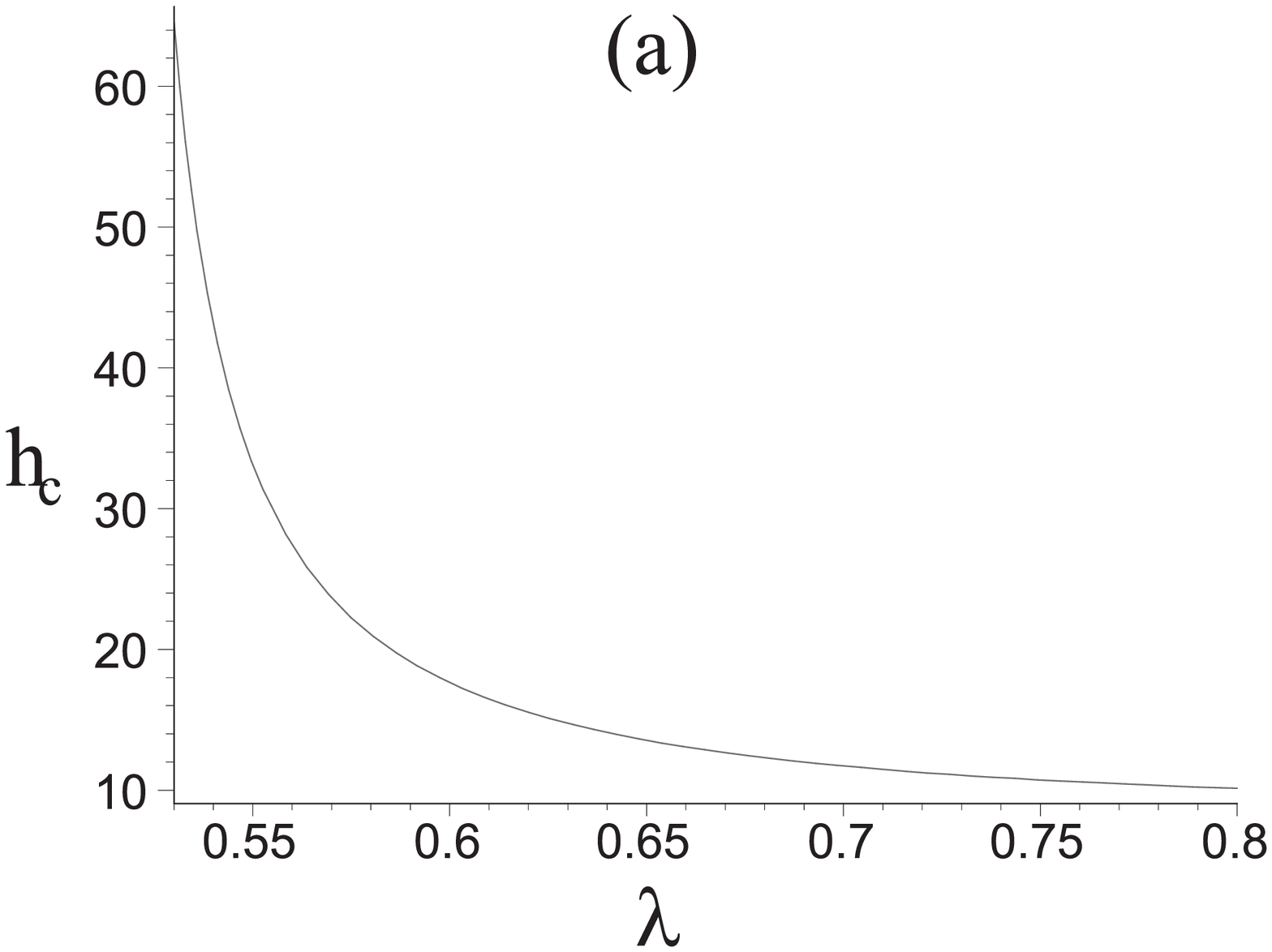}}} \par}
{\par\centering
\resizebox*{8cm}{8cm}{\rotatebox{0}{\includegraphics{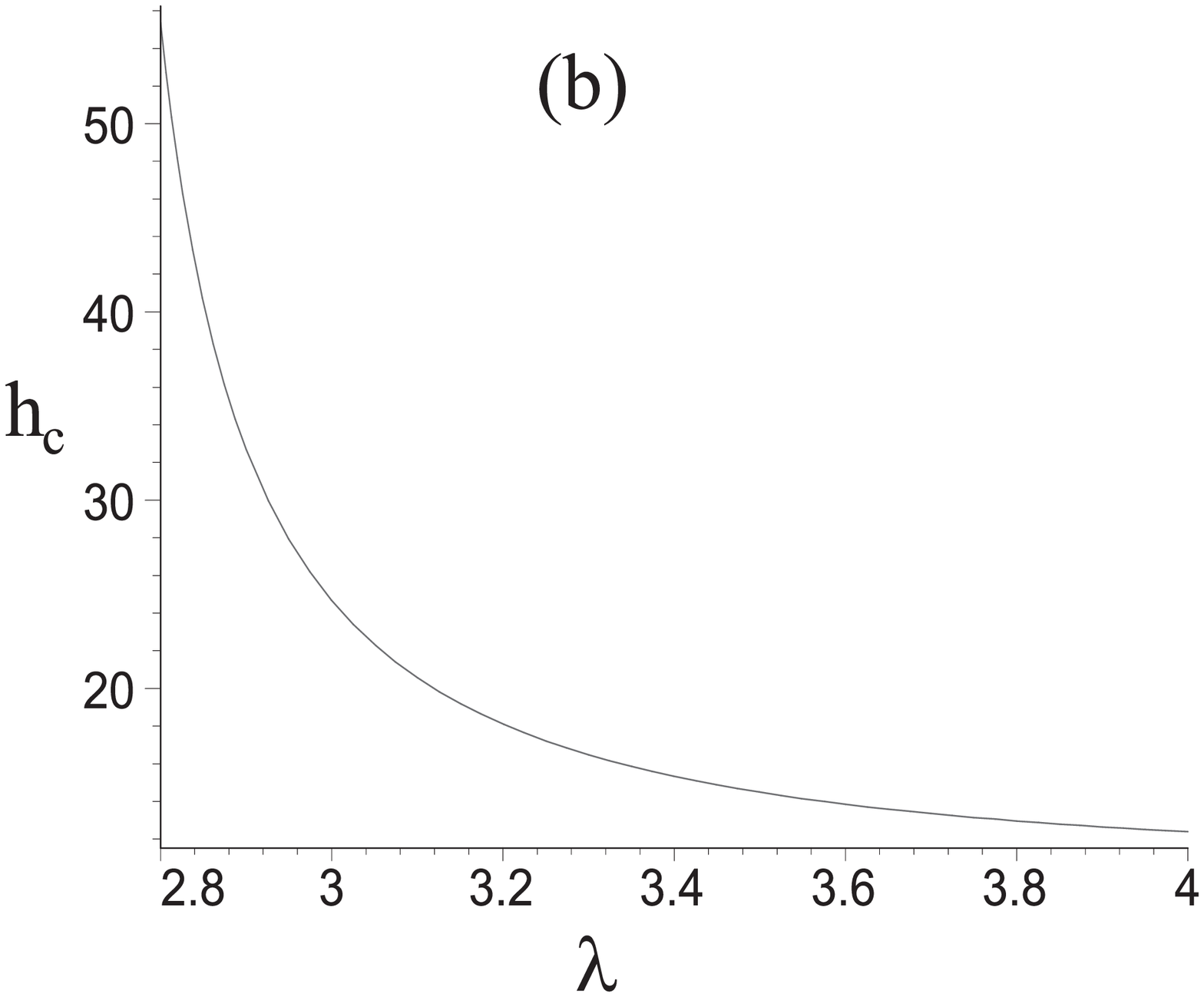}}} \par}
\caption{Critical field $h_c$ as a function of $\lambda$.
Both diagrams correspond to the case of model 1 and the choice of
a finite thickness of the layer $l=1000$.
In Fig.~(a) the lattice model has been used, and in Fig.~(b) the
Carnahan-Startling model. A critical field exists only when $\lambda$
is above a minimum value, which is $0.57$ in the case of the lattice
model, and $2.68$ in the case of the Carnahan-Starling model.
Notice that the critical field $h_c$ tends to a finite limiting value
at infinite value of $\lambda$, which is $9.1$ in the case of the
lattice model, and $11.32$ for the Carnahan-Startling model.}
\label{Fig:lambda}
\end{figure}

For a given value of the dipolar interaction parameter $\lambda$ and the
thickness $l$, the critical point is characterized by a critical
 volume fraction $\ph_c$, and
a critical field $h_c$. In model 1, $\ph_c$ depends on
the entropy only, and is thus independent of the
magnetic field, of $\lambda$ and of $l$.
The critical field $h_c$ on the other hand, does depend on the value
of $\lambda$ and $l$.
In Fig.~\ref{Fig:lambda}, the evolution of $h_c$ in model 1 is
 shown as a function of $\lambda$ for the lattice model in (a)
and for the Carnahan-Starling model in (b).
Both figures correspond to the choice of
a finite thickness of the layer $l=1000$.
The critical field $h_c$ changes slightly as a function of the thickness,
in this regime of large thickness where $q^* l \gg 1$.
A critical field exists only when $\lambda$
is above a minimum value, which is $0.57$ in the case of the lattice
model, and $2.68$ in the case of the Carnahan-Starling model.
If $\lambda$ is smaller than these limits,
there are no equilibrium modulated phases.
Notice also that the critical field $h_c$ tends to a finite limiting value
when $\lambda$ is arbitrary large, which is $9.1$ in the case of the
lattice model, and $11.32$ for the Carnahan-Startling model.
These lower bounds on the critical field are not zero,
since there can be no modulated phases at zero field
in this model as noted before.

\begin{figure}
{\par\centering
\resizebox*{8cm}{8cm}{\rotatebox{0}{\includegraphics{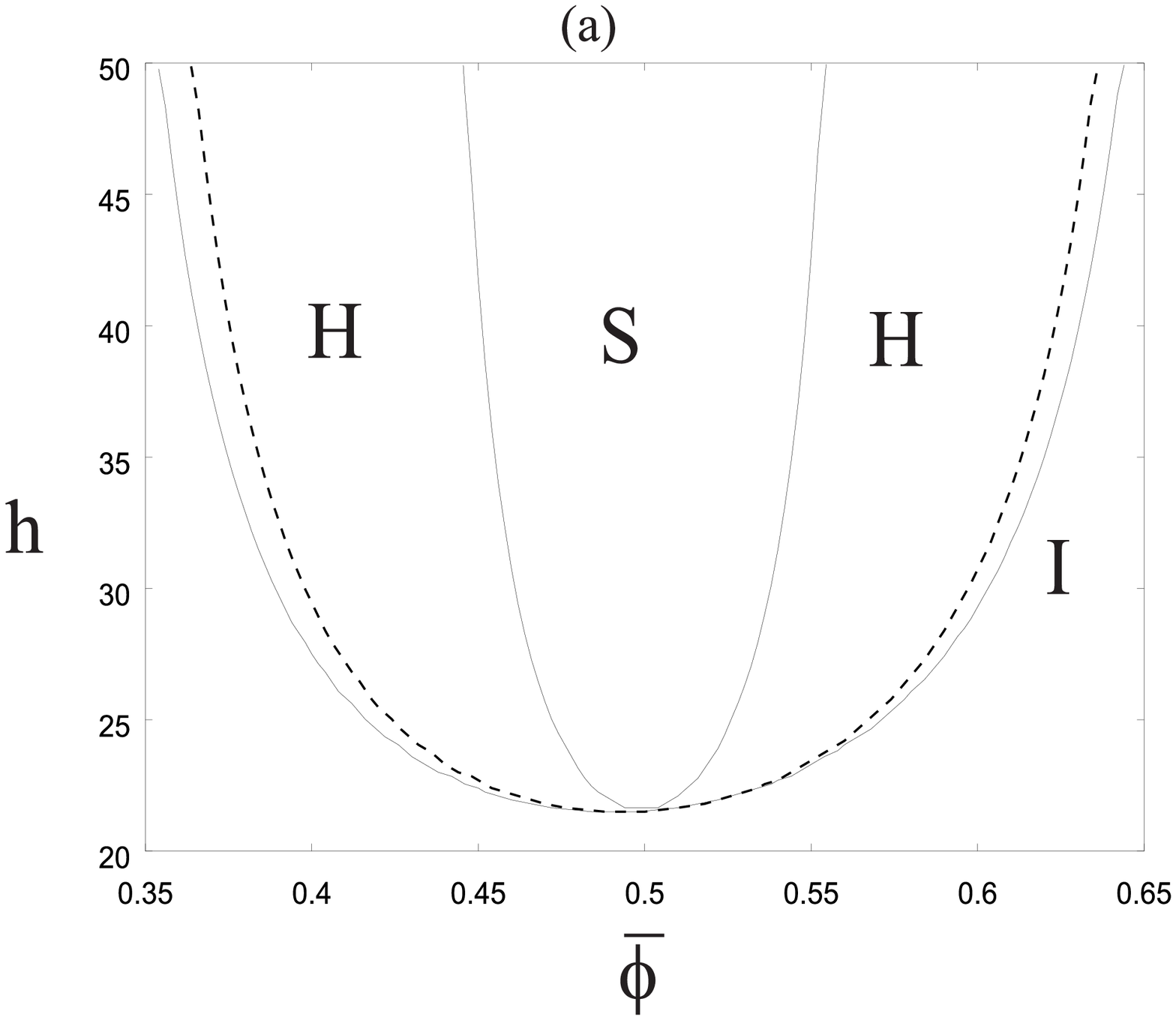}}} \par}
{\par\centering
\resizebox*{8cm}{8cm}{\rotatebox{0}{\includegraphics{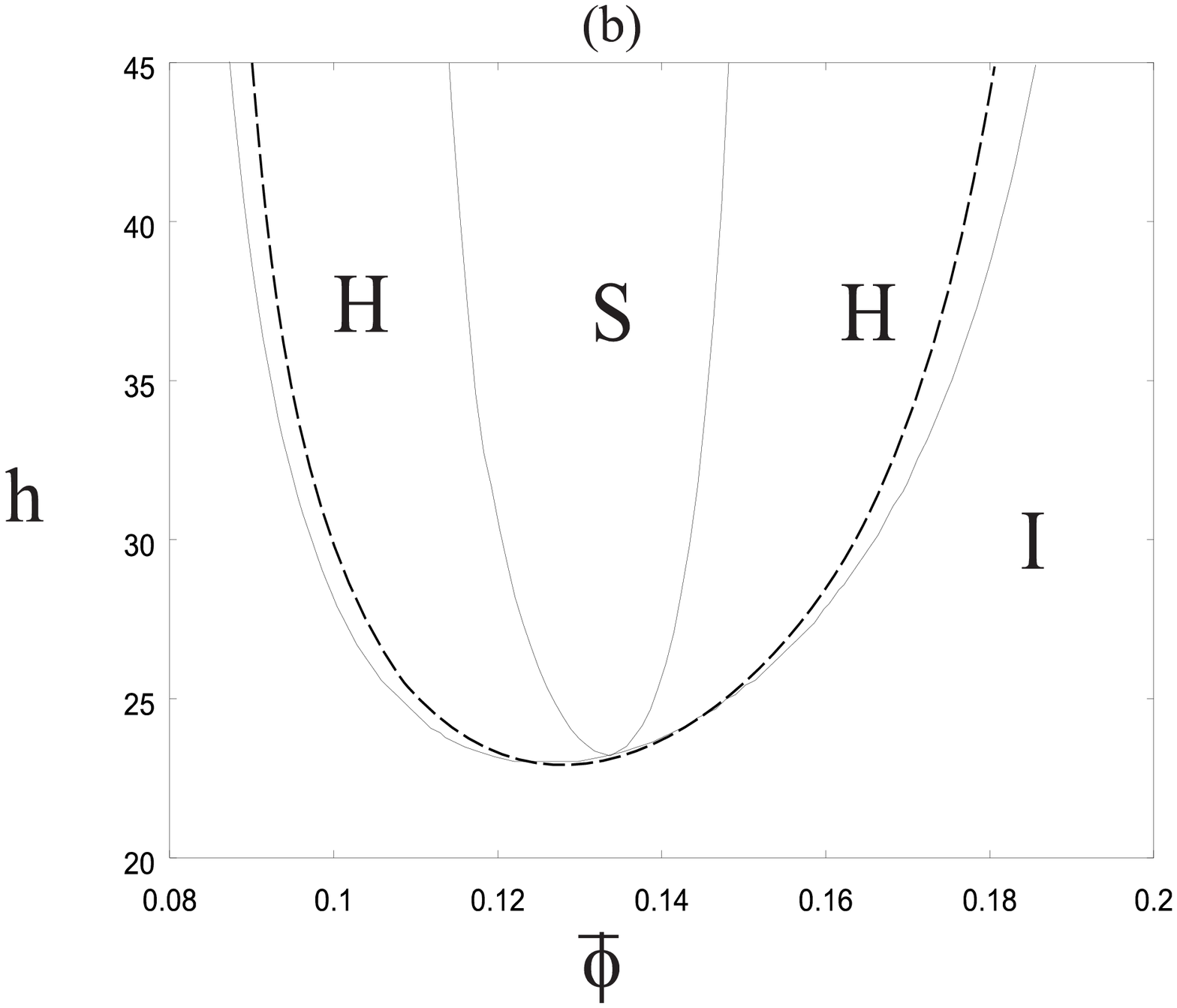}}} \par}
\caption{Phase diagram in the ($h$,$\ph$) plane where $h$ is the
dimensionless magnetic
field and $\ph$ is the average volume fraction of the
ferrofluid. S stands for stripe phase,
H for hexagonal phase and I for uniform phase. Solid lines represent
coexistence lines and the dashed
line is the spinodal. (a) This diagram was obtained for
$\lambda=0.578$, $l=1000$ with model 1 and using the
entropy of a gas on a lattice of Eq.~(\ref{eq:entropy}). (b) This diagram
was obtained for
$\lambda=3$, $l=1000$ with model 2 and using the entropy of  a fluid following
Carnahan-Starling equation of Eq.~(\ref{eq:entropyCS}).}
\label{Fig:1}
\end{figure}

Figure \ref{Fig:1} presents phase diagrams obtained by minimizing the
free energy for the different phases. In these diagrams, the spinodal
line $J(q^*)=0$ is dashed, and the coexistence lines are solid.
Figure \ref{Fig:1}a is the phase diagram for model 1 for $\lambda = 0.578$ and
$l = 1000$ of a system with the lattice-gas entropy
of Eq.~(\ref{eq:entropy}), whereas Fig.\ \ref{Fig:1}b is the phase
diagram for model 2, $\lambda=3$ and $l = 1000$ of a system with the
entropy of a Carnahan-Startling fluid \cite{carnahan}, defined
in Eq.~(\ref{eq:entropyCS}).
In the vicinity of the critical point all phases are present: the
uniform phase (I) (liquid on one side and gas on the other side),
the hexagonal phase (H) (direct hexagonal on one side  and
inverted hexagonal on the other) and the stripe phase (S). The
diagram \ref{Fig:1}a has a lot in common with the one  obtained
by Andelman {\it et al.} \cite{andelman} for Langmuir films and by Cebers
for ferrofluids \cite{cebers2},  with some important differences.
In contrast to the phase diagram of Andelman {\it et al.}, we find
that at high values of the magnetic field (which corresponds to
the magnitude of the electrostatic dipolar interaction in their case),  a ferrofluid
of volume fraction close to $\ph_c=0.5$ has always periodic order.
The disappearance of the stripe and hexagonal phases in the phase diagram of
Andelman {\it et al}. was due to the breakdown of the expansion of the
free energy away from the critical point. In
agreement with predictions by Halsey \cite{halsey}  and Cebers
\cite{cebers}, we find that the stable  ordered phase of a ferrofluid
at low concentration should be the hexagonal phase at low magnetic
field, and the stripe phase at higher magnetic field.

Figure \ref{Fig:1}b shows the phase diagram based on the
Carnahan-Starling description of a liquid of hard spheres, which is
more accurate  than the lattice gas or the Van der Waals models.
For this figure, a value of $\lambda=3$
was chosen. This is the estimated value for a monodisperse suspension
of magnetite particles with $M_s=446$ kAm$^{-1}$, $a=7.4$ nm
and $\delta=1$ nm  at room temperature. For these particles, $h$
is equal to one for a field of $52$ Gauss. The dimensionless length $l$
is estimated to be $1000$, which corresponds to a modulation period of
$1 \mu$m, for $L=40 \mu$m and  a magnetic field $H=300$ Gauss.
This magnetic field is the theoretical critical magnetic field
for microphase separation at a volume fraction of about $1\%$.
As $\lambda$ is increased further, the phase diagram of figures
\ref{Fig:1}a and \ref{Fig:1}b are shifted to lower fields and the modulated phases
extend further away from $\ph_c$.
Quantitative comparison between theory and experiments
require an estimate of $\lambda$, which
presuppose precise determination of particle size and low
polydispersity, as $\lambda$ is proportional to
 the volume of the magnetic particles.

In our model, we have found that the hexagonal phase can coexist with
the uniform phase and the hexagonal phase can coexist with the stripe phase, but
the stripe phase cannot coexist with the uniform phase except at
the critical point. The size of the coexistence regions of H+I and H+S
have been found in our calculations to be small for both the lattice model and
the Carnahan-Startling model (typically of the order of $0.01 \%$
to $0.1 \%$ in volume fraction in figures
\ref{Fig:1}a and \ref{Fig:1}b). For higher values of $\lambda$ however,
the width of the coexistence region of H+S becomes larger.
In contrast, Refs.~\cite{andelman} and \cite{cebers2} find rather large
coexistence regions for both H+S and H+I,
whose origin can be traced to the use of a power-law expansion
of the free energy near a critical point in terms of the uniform part of the particle volume
fraction rather than the full free energy in terms of this variable.

\section{Conclusion}

In this paper, we have presented a picture of the microphase
separation and the formation of ordered phases in ferrofluids under
a magnetic field using mean-field theory and a model of hard  spheres
for the non-dipolar interaction. Within these hypotheses, we have
shown that the attractive part  of the dipolar interaction due to
the local field is responsible for the microphase separation. In
our model, this microphase separation is not possible at zero
magnetic field however large the value of $\lambda$, in agreement
with numerical simulations on the dipolar hard sphere liquid by
Stevens {\it et al.} \cite{stevens,stevens2}. Of course this
conclusion would be changed  if a sufficiently strong isotropic
attraction was added, as in the model of Sano {\it et al.} for
instance \cite{doi}.

We have introduced a theory for phase transitions in pure ferrofluids
based on two order parameters.
At sufficiently high field, where the fluctuations of the magnetic
moment of the particles are small, an effective theory based only on the
volume fraction as an order parameter may be constructed. At infinite
magnetic field, the effective theory is identical to the theory based
on a single order parameter.
We have compared our approach with the work of Cebers \cite{cebers,cebers2}
and found essentially good agreement, with some differences which
have been discussed.

 In order to apply our model to real
 ferrofluids,  some knowledge of the stabilization interaction
 in ferrofluids due to the surfactant is needed. The modeling
 of this non-dipolar part of the interaction  might require more than
 a repulsive hard core, and this will modify the condition of
 microphase separation and the characteristic length of the
 ordered phases. Thermodynamic measurements found evidence for a
 critical liquid-gas transition in ferrofluids, but the precise form
 of the non-dipolar interaction is still not clear in these
 experiments \cite{dubois}. Once the details of this interaction
 are known, the next step towards  a better comparison with
 experiments will introduce polydispersity in the model, which is
 of importance in the context of microphase separation.\\

The authors gratefully acknowledge stimulating discussions with
A. G. Yodh, M. Islam. We thank A. Cebers for a careful reading of the manuscript.
 This work was supported in part by the
MRSEC program under grant NSF DMR00-79909. D. Lacoste received support
by a grant from the French Ministry of Foreign Affairs.

\appendix
\section{Derivation of the function $\fm(m)$}
\label{a:fm}
Let $\fm(m)$ be the free energy per magnetized particle to produce
a magnetic moment $m$. The one particle partition function is
\begin{equation}
Z(h)=\frac{1}{4\pi} \int d\Omega \exp(h\cos \theta)=\frac{\sinh(h)}{h},
\end{equation}
where $h$ has been defined in Eq.~(\ref{eq:defh}).
We define $m$ to be the angular average of the magnetic moment
 over all possible orientations, so that
$m={\mathcal{L}}(h)$.
The free energy associated with the partition function $Z(h)$ is
$g(h)=-k_B T \log (Z(h))$. The function $\fm(m)$ is the Legendre transform
of $g(h)$ with respect to $m$: $\fm(m)=g(h)+mh$. This implies
\begin{equation}
\label{eq:fm2}
\fm(m)=m {\mathcal{L}}^{-1} (m)
 -\log\left(\frac{\sinh \left({\mathcal{L}}^{-1}(m)\right)}
{{\mathcal{L}}^{-1}(m)}\right),
\end{equation}
which is the result of Eq.~(\ref{eq:fm}).
Therefore by construction $\fm(m)$ has the property that
$\partial \fm / \partial m=h={\mathcal{L}}^{-1}(m)$.
Close to $m=0$ the following Taylor expansion is useful
\begin{equation}
\label{eq:taylor}
\fm(m)=\frac{3}{2}m^2+\frac{9}{20}m^4+o(m^6).
\end{equation}
Widom {\it et al.} have used the complete power series of the function
$\fm(m)$ \cite{zhang}.
In general $m$ is not close to $0$, and Eq.~(\ref{eq:taylor}) can not
be used but fortunately it is possible to calculate all the
 derivatives of $\fm(m)$ analytically:
For instance,
\begin{equation}
\fm''(m)=\frac{\partial {\mathcal{L}}^{-1}(m) }{m}=
 \frac{-{\mathcal{L}}^{-1}(m)^2}{-{\mathcal{L}}^{-1}(m)^2+{\mathcal{L}}^{-1}(m)^2
 \coth^2({\mathcal{L}}^{-1}(m))-1}.
\end{equation}
In the limit where $m \rightarrow 1$ which corresponds to complete alignment of
the magnetic moment in the field, it is interesting to note that
$h \simeq 1/(1-m)$ and $\fm'' \simeq 1/(1-m)^2$.

\section{Effective free energy of the stripe and hexagonal phases}
\label{a:eff}
In this appendix, we give the expression of the free energy of
the hexagonal and stripe phase as function of $m_q$ and
$\phi_q$, which are the amplitude of the spatial modulation of the
two order parameters $\phi(\mathbf{r})$ and $m(\mathbf{r})$.
The results take a simple form when two assumptions are made:
it is assumed that the coefficient of the $m_q^2$ term $r_{11}$ is strictly
positive, and that the spatial modulation of
 $\phi(\mathbf{r})$ and $m(\mathbf{r})$ are in phase with each other.
With these assumptions, we derive the effective theory for $\phi$ only,
 when the fluctuations of $m$ have been integrated.
Up to fourth order in $\phi_q$ and third order in $m_q$,
the free energy of the stripe phase is
\begin{equation}
\label{fs2}
f_s=f_{\rm iso} + \frac{1}{2}r_{11} m_{q}^2+\frac{1}{2}r_{22} \phi_{q}^2+
r_{12}m_{q} \phi_{q}+u_1 \phi_{q}m_{q}^3+u_2 m_{q}^2 \phi_{q}^2+u_3 \phi_{q}^4.
\end{equation}
The coefficients of the quadratic part have already been defined in
Eq.~(\ref{eq:det}), the other coefficients are
\begin{eqnarray}
u_1=\frac{1}{16}\fm'''( {\overline m}) \\
u_2=3\lambda g(0)+\frac{3}{2}\lambda g(2q^*) \\
u_3=\frac{1}{64}s^{(4)}(\ph).
\end{eqnarray}
Minimizing $f_s$ with respect to $m_q$ and
 reporting into the result into Eq.~(\ref{fs2}),
one obtains the free energy of Eq.~(\ref{fs}),
which contains the renormalized coefficients $r$ and $u_s$.
The coefficient $r$ has been defined in Eq.~(\ref{eq:renorm}), and
$u_s$ is
\begin{equation}
\label{eq:us}
u_s= u_2 t^2 -u_1 t^3 +u_3,
\end{equation}
with $t=r_{12}/r_{11}$.
The free energy at its minimum is $(f_s)_{min}=-4r^2/u_s$.

For the hexagonal phase, the same procedure results
in the free energy
\begin{eqnarray}
\label{fh2}
f_h=f_{\rm iso} + \frac{3}{2}r_{11} m_{q}^2+\frac{3}{2}r_{22} \phi_{q}^2+
3 r_{12}m_{q} \phi_{q}+15 u_1 \phi_{q}m_{q}^3+\tilde{u_2} m_{q}^2 \phi_{q}^2+15 u_3 \phi_{q}^4 \\
+v_1 m_q^2 \phi_q + v_2 m_q \phi_q^2 + v_3 m_q^3 + v_4 \phi_q^4, \nonumber
\end{eqnarray}
with
\begin{eqnarray}
\tilde{u_2}=18 \lambda \left( 3g(0) + g(\sqrt{3}q^*) +
\frac{1}{4} g(2q^*) + g(q^*) \right) \\
v_1=\frac{3}{4}\fm'''( {\overline m}) +36\lambda \ph g(q^*) \\
v_2=36 \lambda \overline{m} g(q^*) \\
v_3= \frac{1}{4} \fm'''( {\overline m}) \\
v_4= \frac{1}{4} s^{(3)}(\ph).
\end{eqnarray}
The renormalized expression of the free energy of the hexagonal phase
has been given in Eq.~(\ref{fhex}) in terms of the renormalized
coefficients $r,v$ and $u_h$, with
\begin{eqnarray}
\label{eq:v}
v=v_4 + v_1 t^2 - v_2 t - v_3 t^3, \\
\label{eq:u_h}
u_h=-15u_1 t^3 +15 u_3 +\frac{2v_1 v_2 t}{3 r_{11}}-\frac{v_2^2}{6r_{11}}
-\frac{3v_3^2 t^3}{2r_{11}} + \frac{2v_3 t^3 v_1}{r_{11}} +t^2 \tilde{u_2}
-\frac{v_3 v_2 t^2}{r_{11}} -\frac{2v_1^2 t^2}{3r_{11}}.
\end{eqnarray}
In the limit $m \rightarrow 1$, the renormalized coefficients $r,u_h,v$
 tend to the value that these coefficients take in the simpler theory where
the only order parameter is $\phi$. Indeed since $t \simeq (m-1)^2$, $r-r_{22}
 \simeq (m-1)^2$, $v-v_4 \simeq (m-1)^2$, $u_h - 15u_3 \simeq (m-1)^2$ and
$u_s - u_3 \simeq (m-1)^3$. Note that Eq.~(\ref{eq:v}) implies that the
critical volume fraction $\ph_c$ which is the solution of the equation
 $v=r=0$ is now dependent on the magnetic field. In the limit of very high
 field, the critical point should be identical to the critical point of
 the theory with $\phi$ only as order parameter.

In general, we have found that the phase diagram constructed from this
 effective theory is not very different from the phase diagram constructed
 with the theory based on a single order parameter. In the
 limit of infinite field, this effective theory becomes identical
to the theory with a single order parameter (model 1).

\begin{thebibliography}{10}
\bibitem{rosensweig} R.~E. Rosensweig,
\newblock {\em Ferrohydrodynamics}.
\newblock Dover, New York, 1997.

\bibitem{bacri1} J-C. Bacri and D.~Salin,
\newblock {\em J. Phys. (Paris), Lett.}, 43:L771--L777, 1982.

\bibitem{hong} C-Y. Hong, I.~J. Jang, H.~E. Horng, C.~J. Hsu, Y.~D.
Yao, and H.~C. Yang,
\newblock {\em J. Appli. Phys.}, 81(8):4275--7, 1997.
\bibitem{wang} H.~Wang, Y.~Zhu, C.~Boyd, W.~Luo, A.~Cebers, and R.~E.
Rosensweig,
\newblock {\em Phys. Rev. Lett.}, 72(12):1929--32, 1994.

\bibitem{flament} C.~Flament, J.-C. Bacri, A.~Cebers, F.~Elias, and
R.~Perzynski,
\newblock {\em Europhys. Lett.}, 34(3):225--30, 1996.

\bibitem{bibette} J.~Liu, E.~M. Lawrence, A.~Wu, M.~L. Ivey, G.~A.
Flores, K.~Javier, J.~Bibette,   and J.~Richard,
\newblock {\em Phys. Rev. Lett.}, 74(14):2828--31, 1995.

\bibitem{andelman} D.~Andelman, F.~Brochard, and J-F. Joanny,
\newblock {\em J. Chem. Phys.}, 86(6):3673--81, 1987.

\bibitem{garel} T.~Garel and S.~Doniah,
\newblock {\em Phys. Rev. B}, 26(1):325--9, 1982.

\bibitem{faber} T.~E. Faber,
\newblock {\em Proceeding of the Royal Society London A}, 248:460--81,
1958.

\bibitem{seul} M.~Seul and R.~Wolfe,
\newblock {\em Phys. Rev. A}, 46(12):7519--33, 1992.

\bibitem{ytreberg} F.~M. Ytreberg and S.~R. McKay,
\newblock {\em Phys. Rev. E}, 61(4):4107--10, 2000.

\bibitem{carnahan} N.~F. Carnahan and K.~E. Starling,
\newblock {\em J. Chem. Phys.}, 51(2):635--63, 1969.

\bibitem{morozov} K.~I. Morozov,
\newblock {\em Magnetohydrodynamics}, 23(1):37--41, 1987.
\bibitem{doi} K.~Sano and M.~Doi,
\newblock {\em J. Phys. Soc. Jap.}, 52(8):2810--5, 1983.

\bibitem{cebers} A.~Cebers,
\newblock {\em Magnetohydrodynamics}, 18(2):137--142, 1982.

\bibitem{cebers2} A.~Cebers,
\newblock {\em Magnetohydrodynamics}, 31(1,2):58--72, 1995;
\newblock {\em Magnitnaya Gidrodinamika}, 35(4):344--363, 1999;
\newblock {\em Proceedings of the Forth International PAMIR
conference on Magnetohydrodynamics}, 18-22:175--180, 2000.

\bibitem{halsey} T.~C. Halsey,
\newblock {\em Phys. Rev. E}, 48(2):R673--5, 1993.

\bibitem{lubensky} P.~M. Chaikin and T.~C. Lubensky,
\newblock {\em Principles of condensed matter physics}.
\newblock Cambridge University Press, New York, 1995.

\bibitem{zhang} H.~Zhang and M.~Widom,
\newblock {\em Phys. Rev. E}, 49(5):R3591--3, 1994.

\bibitem{degennes} P.~G. de~Gennes and P.~A. Pincus,
\newblock {\em Phys. Kondens. Materie}, 11:188--198, 1970.

\bibitem{Ivey} M.~Ivey, J.~Liu, Y.~Zhu, and S.~Cutillas,
\newblock {\em Phys. Rev. E}, 63:011403, 2000.

\bibitem{stevens} M.~J. Stevens and G.~S. Grest, %
\newblock {\em Phys. Rev. Lett.}, 72(23):3686--9, 1994.

\bibitem{stevens2} M.~J. Stevens and G.~S. Grest,
\newblock {\em Phys. Rev. E}, 51(6):5962--75, 1995.

\bibitem{dubois} E.~Dubois, V.~Cabuil, F.~Bou\'e, and R.~Perzynski,
\newblock {\em J. of Chem. Phys.}, 111(15):7147--60, 1999.

\bibitem{brazovskii} S.~A.~Brazovskii,
\newblock {\em Sov. Phys. JETP}, 41(1):85--89, 1975.

\end{thebibliography}

\end{document}